\documentclass[journal=accacs,manuscript=article,layout=singlecolumn]{achemso}
\setkeys{acs}{usetitle=true,maxauthors=99}


\usepackage{amssymb}
\usepackage{amsmath}
\usepackage{graphicx}
\usepackage{dcolumn}
\usepackage{bm}
\usepackage[utf8]{inputenc}
\usepackage[T1]{fontenc}
\usepackage{mathptmx}
\usepackage{etoolbox}
\usepackage{doi}

\usepackage{xcolor}
\usepackage{siunitx} 
\usepackage{chemformula}
\usepackage{comment}
\usepackage[shortlabels]{enumitem}
\usepackage[english]{babel}
\usepackage[draft=false]{microtype}

\newcommand{\expP}[1]{{#1 ^ \circ}}
\newcommand{\expTS}[1]{{#1 ^ \ddagger}}

\newcommand{\esP}[1]{{#1 ^ \circledast}}
\newcommand{\stP}[1]{{#1 ^ {\ast}}}

\renewcommand{\vec}{\overrightarrow}
\newcommand{\mat}{\underline}
\let\oldnabla\nabla
\renewcommand{\nabla}{\vec{\oldnabla}}

\title{A Theoretical Investigation of the Grand- and the Canonical Potential Energy Surface: The Interplay between Electronic and Geometric Response at Electrified Interfaces}

\author{Simeon D. Beinlich}
\email{* beinlich@fhi.mpg.de}
\affiliation{Fritz-Haber-Institut der Max-Planck-Gesellschaft, Faradayweg 4-6, 14195 Berlin, Germany}
\alsoaffiliation{Technical University of Munich, Lichtenbergstr. 4, 85747 Garching, Germany}

\author{Georg Kastlunger}
\affiliation{Technical University of Denmark, Fysikvej 311, 2800 Kongens Lyngby, Denmark}

\author{Karsten Reuter}
\affiliation{Fritz-Haber-Institut der Max-Planck-Gesellschaft, Faradayweg 4-6, 14195 Berlin, Germany}

\author{Nicolas G. H{\"o}rmann}
\affiliation{Fritz-Haber-Institut der Max-Planck-Gesellschaft, Faradayweg 4-6, 14195 Berlin, Germany}


\begin{document}
    
\begin{abstract}
How does an electrochemical interface respond to changes in the electrode potential?
How does the response affect the key properties of the system -- energetics, excess charge, capacitance?
Essential questions key to ab-initio simulations of electrochemical systems, which we address in this work on the basis of a rigorous mathematical evaluation of the interfacial energetics at constant applied potential.
By explicitly taking into account the configurational and electronic degrees of freedom we derive important statements about stationary points in the electronically grand canonical ensemble.
We analyze their geometric response to changes in electrode potential and show that it can be mapped identically onto an additional contribution to the system's capacitance.
We draw similar conclusions for the constant charge ensemble which equally allows to assess the respective stationary points.
Our analysis of the relation between the canonical and grand canonical energetics reveals, however, one key difference between both ensembles.
While the constant potential ensemble yields in general positive capacitances at local minima, the capacitance of local minima in the constant charge ensemble might become negative.
We trace back this feature to the possibility of character switching of stationary points when switching between the ensembles causing the differences in the response to perturbations. 

Our systematical analysis not only provides a detailed qualitative and quantitative understanding of the interplay between electronic and configurational degrees of freedom and their contributions to the energetics of electrified interfaces but also highlights the similarities and subtle dissimilarities between the canonical and grand canonical description of the electronic degrees of freedom, which is crucial for a better understanding of theoretical calculations with and without potentiostat.
\end{abstract}

    \clearpage
    \section{Introduction}
    The behavior of electrified interfaces under an applied potential is central to many fields of study, from energy storage systems to electronic devices. A key parameter in this regard is the system's capacitance, a metric that characterizes the system's response to voltage changes. It can be assessed by measuring the change in electronic excess charge $\mathrm{d} q$ induced by a small change in electrode potential $\mathrm{d} U$ via
    \begin{align}
        C = \frac{\mathrm{d} q }{\mathrm{d} U}\nonumber\quad,
    \end{align}
    where $U$ can be directly related to the electrons' electrochemical potential. As a result, the quantity $q$ can be traced back to a grand canonical (GC) potential energy $\mathcal{E}$, where by definition $q = -\tfrac{\mathrm{d} \mathcal{E}}{\mathrm{d} U}$.
    
    In practice $\mathcal{E}$ is a complex function of composition, configuration, and other independent degrees of freedom making it a function in multiple dimensions. In atomistic calculations with fixed composition containing $N$ atoms, the potential energy surface (PES), the space spanned by the degrees of freedom of $\mathcal{E}$, is a function of $U$ and $3N$ spatial coordinates.
    A typical situation where this can be recognized is when studying a certain state of the interface, e.g. a molecule adsorbed on the surface, at constant potential conditions. At a specified applied potential $U$, the system will adapt its geometry to minimize $\mathcal{E}$. A small change in potential will modify $\mathcal{E}$ and as a result shift the geometry to the new minimum of $\mathcal{E}$ at the new potential.
    
    This paper explores $\mathcal{E}$ as a function of the potential and geometric degrees of freedom, revealing the electronic and geometric response to a change in potential that determines e.g. the capacitance $C$. This allows to make several important statements about the grand canonical energetics and the capacitance of stationary points. In addition, we perform a similar analysis in the canonical ensemble and dissect the intricate relationship between the grand canonical (constant potential) PES, and the canonical (constant charge) PES, which is the intrinsic energy landscape of common ab-initio density functional theory (DFT) calculations.
    
    This work paves the way for a deeper understanding of the response of electrified interfaces to an applied potential.
    
    \section{Theory}
    The focus of this work is to understand the behaviour of stationary points on both the grand canonical and the canonical potential energy surfaces (gcPES and cPES) and their response to changes in the respective independent variables -- the applied electrode potential $U$ in case of the gcPES and the excess charge $q$ in case of the cPES respectively.
    Note, that grand canonical and canonical relate in the present work only to the description of the electronic degrees of freedom and not of the atoms in the system.
    This possibility and necessity to treat electrons and ions independently is specific to electrochemical systems, and as we leave the (ionic) composition unaltered in all cases, we use the term PES.
    
    The analysis is largely based on considering the PES as an explicit function of $\vec{r}$  -- the $3N$ spatial degrees of freedom of a system containing $N$ atoms --  and the additional electronic degree of freedom $U$ or $q$, respectively.
    
    \subsection{Stationary points on the grand canonical PES}
    We start with evaluating the grand canonical PES $\mathcal{E}(U, \vec{r})$.
    For this, we define the excess charge $q$ and the electronic capacitance $C_{\rm el}$ as the first and second derivatives in the electronic degree of freedom $U$ (cf. Ref.~\citenum{bonnet2012first, hormann2019grand, hormann2020electrosorption, lindgren2022electrochemistry}), the Force $\vec{\mathcal{F}}$ and the Hessian $\mat{\mathcal{H}}$ as those in the geometric degrees with the respective signs as listed in Table~\ref{tab:derivatives} (cf. Ref.~\citenum{lindgren2022electrochemistry}).
    The mixed second derivatives, the configurational gradient of the excess charge $\nabla q$ and the derivative of the force with respect to electrode potential $\tfrac{\partial \vec{\mathcal{F}}}{\partial U}$, are identical due to the interchangeability of second derivatives (Schwarz's theorem, cf. Ref.~\citenum{vijay2022force}):
    \begin{align}
        \left.\tfrac{\partial q}{\partial r_i}\right|_U 
        = \left.\tfrac{\partial }{\partial r_i} \left(- \left.\tfrac{\partial \mathcal{E}}{\partial U}\right|_r \right)\right|_U
        = \left.\tfrac{\partial }{\partial U} \left(-\left.\tfrac{\partial \mathcal{E}}{\partial r_i}\right|_U \right)\right|_r 
        = \left.\tfrac{\partial \mathcal{F}_i}{\partial U}\right|_r\label{eq:grad_q_dF_dU}\quad.
    \end{align}
    Note, that each of these grand canonical quantities has to be considered a function of the independent variables $(U, \vec{r})$, electrode potential and system geometry, respectively.
    \begin{table}[t]
        \centering
        \begin{tabular}{c|c|l}
            \hline
            derivative                                                  &                                quantity                       &          description           \\ \hline
            $\tfrac{\partial \mathcal{E}}{\partial U}$                  &                                  $-q$                         &               excess charge    \\
            $\tfrac{\partial^2 \mathcal{E}}{\partial U^2}$              &                          $- C_{{\rm el}}$                     &    electronic capacitance      \\
            $\tfrac{\partial \mathcal{E}}{\partial r_i} $               &                         $- \mathcal{F}_i$                     &    grand canonical forces      \\
            $\tfrac{\partial^2 \mathcal{E}}{\partial r_j \partial r_i}$ &                              $\mathcal{H}_{ij}$               &    grand canonical Hessian     \\
            $\tfrac{\partial^2 \mathcal{E}}{\partial r_i \partial U}$   &           $-\tfrac{\partial q}{\partial r_i}$                 &   variation of excess charge caused by spatial displacements \\ 
            $=\frac{\partial^2 \mathcal{E}}{\partial U \partial r_i}$   &           $ -\frac{\partial \mathcal{F}_i}{\partial U} $      & variation of forces caused by changes in potential           \\ \hline
        \end{tabular}
        \caption{First and second derivatives of the $3N+1$ dimensional grand canonical energy $\mathcal{E}(U, \vec{r})$.
            Note that each quantity as well depends on the applied potential and the geometry of the system $(U, \vec{r})$.
            The mixed second derivatives are identical due to Schwarz's theorem (see Eq.~\ref{eq:grad_q_dF_dU}).}
        \label{tab:derivatives}
    \end{table}

    A stationary point $\stP{\vec{r}}$ is defined as a point on the gcPES where all force components vanish: $\mathcal{F}_i(U, \stP{\vec{r}}) = 0$.
    However, a change in applied electrode potential $U$ will again exert forces on the system requiring it to adapt its geometry to restore the stationarity.
    We hence have to consider $\stP{\vec{r}}$ to be a function of electrode potential $\stP{\vec{r}}(U)$.
    As an example, think of a molecule adsorbed on the surface that caused by an electrode potential change adapts its geometry until all forces acting on it again vanish.
    The specific geometry $\stP{\vec{r}}= \stP{\vec{r}}(U)$ of the stationary point at any potential $U$ is uniquely defined by the stationarity criterion
    \begin{align}
        \stP{\vec{\mathcal{F}}}(U) = \vec{\mathcal{F}}(U, \stP{\vec{r}}) = 0 = \rm const\quad .\label{eq:stat_criterion}
    \end{align}
    Note, that we can treat $\stP{\vec{r}}(U)$ as a smooth, continuous $3N$-dimensional path along which the stationary point moves when we gradually increase the electrode potential $U$.
    Evidently, it is possible that a local minimum only exists in a specific potential window and might become unstable outside of it.
    However, in this case the system falls into a neighboring local minimum that again is stable at the respective potential and that again follows a smooth path.
    Without loss of generality we can, hence, consider $\stP{\vec{r}}= \stP{\vec{r}}(U)$ to be continuous and differentiable, and we can identify the path as a well defined \textit{stationary state} of the system.
    The grand canonical energy of such a stationary state $\stP{\mathcal{E}}$ in dependence of potential $U$ is uniquely defined via the gcPES as $\stP{\mathcal{E}}(U)=\mathcal{E}(U, \stP{\vec{r}}(U))$.\footnote{As an important remark on notation: throughout this work, we continue indicating quantities $X$ evaluated at stationary points $\stP{\vec{r}}$ as $\stP{X}$, i.e. we shorthand write $\stP{X}(U)=X(U, \stP{\vec{r}}(U))$, emphasizing that we evaluate the quantity at the point, i.e. by writing $\stP{\nabla X}$ we mean $\stP{\left(\nabla X \right)} = \left(\nabla X \right)(U, \stP{\vec{r}}(U))$, similar for partial derivatives $\stP{\tfrac{\partial {X}}{\partial U}}=\stP{\left(\tfrac{\partial {X}}{\partial U}\right)} = \left(\tfrac{\partial {X}}{\partial U}\right)(U, \stP{\vec{r}}(U))$. Total derivatives, however, do act on the composed function, e.g. $\tfrac{\mathrm{d} \stP{{X}}}{\mathrm{d} U} = \tfrac{\mathrm{d}}{\mathrm{d} U} \stP{{X}}(U) = \tfrac{\mathrm{d}}{\mathrm{d} U} \left({X}(U, \stP{\vec{r}}(U))\right) = \tfrac{\partial \stP{{X}}}{\partial U} + \sum_i \tfrac{\partial \stP{{X}}}{\partial r_i}\cdot \tfrac{\mathrm{d} \stP{r_i}}{\mathrm{d} U}$ (where we also acknowledged that $\stP{\vec{r}}$ is a pure function of $U$, i.e. the total derivative is identical to the partial derivative).}
    
    From these very general considerations we can already make a first important statement: the total derivative of the grand canonical energy $\stP{\mathcal{E}}(U)$ with respect to potential of a stationary state is identical to its partial derivative, the excess charge $q$, at given potential $U$:
    \begin{align}
        \tfrac{\mathrm{d} \stP{\mathcal{E}}(U)}{\mathrm{d} U} = \tfrac{\mathrm{d}}{\mathrm{d} U} \mathcal{E}(U, \stP{\vec{r}}(U))  = \stP{\tfrac{\partial \mathcal{E}}{\partial U}} + \underbrace{\sum_i \stP{\tfrac{\partial \mathcal{E}}{\partial r_i}}\cdot \tfrac{\mathrm{d} \stP{r_i}}{\mathrm{d} U}}_{=0} = \stP{\tfrac{\partial \mathcal{E}}{\partial U}} = -q(U, \stP{\vec{r}}(U)) = -\stP{q}(U) \label{eq:dE_dU_gc_eneral}\quad,
    \end{align}
    where we used that $\stP{\tfrac{\partial \mathcal{E}}{\partial r_i}} = -\stP{\mathcal{F}_i}=0$ for stationary points (Eq.~\ref{eq:stat_criterion}).
    
    Furthermore, we can derive the general potential-dependence of the path $\stP{\vec{r}}$ -- again using the stationarity condition (Eq.~\ref{eq:stat_criterion}) and taking into account that the forces vanish along the entire path $\stP{\vec{\mathcal{F}}}(U) = 0 = \rm const$, i.e. $\tfrac{\mathrm{d}\stP{\vec{\mathcal{F}}}}{ \mathrm{d} U} = 0$:
    \begin{align}
        0\stackrel{!}{=} \tfrac{\mathrm{d}\stP{\mathcal{F}_i}}{ \mathrm{d} U} = \tfrac{\mathrm{d}}{ \mathrm{d} U} \mathcal{F}_i(U, \stP{\vec{r}}(U))= \left(\stP{\tfrac{\partial \mathcal{F}_i}{\partial U}} + \sum_j \stP{\tfrac{\partial \mathcal{F}_i}{\partial r_j}}  \tfrac{\mathrm{d} \stP{r_j}}{\mathrm{d} U}\right) = \left(\stP{\tfrac{\partial \mathcal{F}_i}{\partial U}} - \sum_j \stP{\mathcal{H}_{i,j}}  \tfrac{\mathrm{d} \stP{r_j}}{\mathrm{d} U}\right)\quad.\nonumber
    \end{align}
    Rearranging  and multiplying from the left with the inverse of the grand canonical Hessian $\mat{\mathcal{H}}^{-1}$ yields:
    \begin{align}
        \sum_j \stP{\mathcal{H}_{i,j}}  \tfrac{\mathrm{d} \stP{r_j}}{\mathrm{d} U} 
        &= \stP{\tfrac{\partial \mathcal{F}_i}{\partial U}}\nonumber\\
        \sum_{j} \underbrace{\sum_i \stP{\mathcal{H}_{k,i}}^{-1} \stP{\mathcal{H}_{i,j}}}_{\delta_{k,j}} \tfrac{\mathrm{d} \stP{r_j}}{\mathrm{d} U} 
        &= \sum_i  \stP{\mathcal{H}_{k,i}}^{-1}  \stP{\tfrac{\partial \mathcal{F}_i}{\partial U}} \nonumber\\
        \frac{\mathrm{d} \stP{r_k}}{\mathrm{d} U}(U) 
        &= \sum_i  \stP{\mathcal{H}_{k,i}}^{-1}  \stP{\tfrac{\partial \mathcal{F}_i}{\partial U}} \quad, \label{eq:dr_dU_general}
    \end{align}
    where we used $\sum_i \stP{\mathcal{H}_{k,i}}^{-1}\stP{\mathcal{H}_{i,j}} = \delta_{k,j}$ with $\delta_{k,j}$ denoting the Kronecker delta.
    While the practical use of this formula might not directly be apparent, it does reveal how the geometry of a stationary point changes with potential: while the forces caused by a change in potential displace the system in the direction of $\tfrac{\partial \vec{\mathcal{F}}}{\partial U}$, the magnitude and also the direction of the displacement is finally determined by the 'chemical' energy landscape that is encoded in the (inverse) Hessian $\mat{\mathcal{H}}^{-1}$.
    
    We can now approach our goal, namely determining the total capacitance, including geometric effects from only these two derived formulas within the constant-potential ensemble. The total capacitance is the negative curvature of $\stP{\mathcal{E}}(U)$ with respect to $U$:
    \begin{align}
        \stP{C_{\rm total}}(U) &= -\tfrac{\mathrm{d}^2 \stP{\mathcal{E}}(U)}{\mathrm{d}U^2} 
        =  \tfrac{\mathrm{d}\stP{q}(U)}{\mathrm{d} U}
        = \tfrac{\mathrm{d}}{\mathrm{d} U}{q}(U, \stP{\vec{r}}(U)) \nonumber\\
        &=   \stP{\tfrac{\partial {q}}{\partial U}} + \sum_i  \stP{\tfrac{\partial {q}}{\partial r_i}}  \tfrac{\mathrm{d} \stP{r_i}}{\mathrm{d} U} 
        =   \stP{C_{\rm el}} + \sum_{i,j}  \stP{\tfrac{\partial {q}}{\partial r_i}}  \stP{\mathcal{H}_{i,j}}^{-1}  \stP{\tfrac{\partial \mathcal{F}_j}{\partial U}} \nonumber\\
        &=  \stP{C_{\rm el}} + \sum_{i,j}  \stP{\tfrac{\partial {q}}{\partial r_i}}  \stP{\mathcal{H}_{i,j}}^{-1}  \stP{\tfrac{\partial {q}}{\partial r_j}} \nonumber\\
        &=  \stP{C_{\rm el}} + \underbrace{ \stP{\nabla q}^T  \stP{\mat{\mathcal{H}}}^{-1} \stP{\nabla q}}_{ \stP{C_{\rm geom}}} \quad, \label{eq:total_cap_fully_analytically_gc}
    \end{align}
    where we used matrix notation and Schwarz's theorem (Eq.~\ref{eq:grad_q_dF_dU}).\footnote{Note that we can define $C_{\rm total}$ and $C_{\rm geom}$ more generally for any point $(U, \vec{r})$ (not only for stationary points $(U, \stP{\vec{r}}(U))$) even though both quantities loose their physical meaning.}
    
    This result is a central finding of our derivations, as it essentially states, that geometric degrees of freedom contribute to the second order potential-dependence of the grand canonical energy of stationary points just as an additional capacitive contribution.
    This expression allows many more fundamental statements, which we discuss in detail below.
    
    First, however, we will show that we can also derive this result in a more illustrative way.
    We do this by approximating the general gcPES $\mathcal{E}(U, \vec{r})$ in the vicinity of stationary points both in the electronic, as well as in the geometric degrees of freedom to second order.
    
    \paragraph{Grand canonical PES in the vicinity of arbitrary points}\label{sec:gen-gc-2ndO-TE}
    First, we derive the generic second order expansion of the grand canonical energy $\mathcal{E}(U, \vec{r})$ defining the $3N+1$ dimensional gcPES around an arbitrary point.
    Using the definitions summarized in Table~\ref{tab:derivatives}, the second order Taylor expansion of $\mathcal{E}(U, \vec{r})$ in the potential $U$ and the $3N$ dimensional coordinate vector $\vec{r}$ around an arbitrary point~$(\expP{U}, \expP{\vec{r}})$ reads:
    \begin{align}
        &{\mathcal{E}}(\expP{U} + \Delta U, \expP{\vec{r}} + \vec{\Delta r}) \nonumber\\
        &= {\expP{\mathcal{E}}} 
        +  \expP{\left(\tfrac{\partial \mathcal E}{\partial U}\right)} \Delta U
        +  \sum_i \expP{\left(\tfrac{\partial \mathcal{E}}{\partial r_i}\right)} \Delta r_i \nonumber\\
        &+ \tfrac{1}{2}\left( \sum_{i,j} \Delta r_j \expP{\left(\tfrac{\partial^2 \mathcal{E}}{\partial r_j \partial r_i}\right)} \Delta r_i\right. 
        + 2\sum_i \expP{\left(\tfrac{\partial^2 \mathcal{E}}{\partial r_i \partial U}\right)} \Delta r_i \, \Delta U 
        +  \left. \expP{\left(\tfrac{\partial^2 \mathcal{E}}{\partial U^2} \right)} \Delta U^2 \right)
        + \mathcal{O}(\Delta U^3, \Delta r ^3) \nonumber\\
        &= {\expP{\mathcal{E}}} 
        - \expP{q} \Delta U  - \sum_i \expP{\mathcal{F}_{i}} \Delta r_i \nonumber\\
        &+ \tfrac{1}{2}\left( \sum_{i,j} \Delta r_j \expP{\mathcal{H}_{i,j}} \Delta r_i
        - 2 \sum_i \expP{\left(\tfrac{\partial q}{\partial r_i}\right)}  \Delta r_i \, \Delta U
        - \expP{C_{\rm el}} \,  \Delta U^2 \right) 
        + \mathcal{O}(\Delta U^3, \Delta r ^3)
        \quad, \label{eq:2ndO_TE_gc}
    \end{align}
    marking all quantities evaluated at the expansion point $(\expP{U}, \expP{\vec{r}})$ with $"\expP{ }"$. This yields an analytical expression for the grand canonical energy that is accurate up to second order around the point $(\expP{U}, \expP{\vec{r}})$. 
    
    \paragraph{Potential-induced geometric shifts of stationary points on the gcPES}
    Now, on this expanded, $(3N +1)$-dimensional parabolic gcPES, where all partial second derivatives are constant per construction, we can evaluate the geometric shift of a stationary point $\stP{\vec{r}}$  analytically.
    Given that the expansion above is known for a stationary point $\esP{\vec{r}}$ at a certain potential $\expP{U}$, we can compute the geometric shift $\Delta\stP{\vec{r}} = \stP{\vec{r}} - \esP{\vec{r}}$ on it that is caused by a change in potential ${\Delta U} = {U} - \expP{U}$. 
    For this, we evaluate the stationarity criterion (Eq.~\ref{eq:stat_criterion}) at the expansion and the target point, i.e. using that both states $(\expP{U},\esP{\vec{r}})$ and $(\expP{U} + {\Delta U}, \esP{\vec{r}} + \stP{\vec{\Delta r}})$ are stationary points where all forces vanish:
    \begin{align}
        0 \stackrel{!}{=} {\mathcal{F}}_i(\expP{U}+{\Delta U}, \esP{\vec{r}} + \stP{\vec{\Delta r}} )
        &=  \underbrace{\esP{{\mathcal{F}}_{i}}}_{=0} + \esP{\left(\tfrac{\partial \mathcal{F}_i}{\partial U}\right)} {\Delta U} - \sum_{j}\esP{\mathcal{H}_{i,j}} \stP{\Delta r_j} + \mathcal{O}(\Delta U^2, \Delta r ^2)\nonumber \\
        \Leftrightarrow \stP{{\Delta r}_{k}}(\Delta U) &=  \sum_i \esP{\mathcal{H}_{k,i}}^{-1} \esP{\left(\tfrac{\partial \mathcal{F}_i}{\partial U}\right)} \Delta U + \mathcal{O}(\Delta U^2)\nonumber\\
        &=  \sum_i \esP{\mathcal{H}_{k,i}}^{-1} \esP{\left(\tfrac{\partial q}{\partial r_i}\right)} \Delta U + \mathcal{O}(\Delta U^2) \quad, \label{eq:delta_r_x_gc}
    \end{align}
    marking a quantity evaluated at the stationary expansion point $(\expP{U}, \esP{\vec{r}})$ as $\esP{X}$. We will use this notation throughout this work.\footnote{As a further clarification on the difference between $\stP{X}$ and $\esP{X}$: $\stP{X}(U)$ essentially is a shortcut for writing $X(U, \stP{\vec{r}}(U))$, i.e. for denoting quantities evaluated at the stationary point at the desired potential. The additional circle in $\esP{X}$ indicates a property evaluated at a stationary expansion point. Essentially, we derive quantities $\stP{Y}$ at a desired potential $U$ based on quantities $\esP{X}$ evaluated at a stationary expansion point $(\expP{U}, \stP{\vec{r}}(\expP{U}))$, i.e. at a different potential $\expP{U}$.} 
    Equation \ref{eq:delta_r_x_gc} yields an alternative formulation of Equation~\ref{eq:dr_dU_general} using the exchanged second mixed derivative (Eq.~\ref{eq:grad_q_dF_dU}).
    
    \paragraph{Grand canonical energy of a stationary point}
    We can now insert the just derived potential-dependent geometric shift of a stationary point $\stP{\vec{\Delta r}}(\Delta U)$ (Eq.~\ref{eq:delta_r_x_gc}) into the second-order expansion of the gcPES (Eq.~\ref{eq:2ndO_TE_gc}) around a known stationary point $(\expP{U}, \esP{\vec{r}})$, eliminating the spatial dependence and returning the potential-dependent
    grand canonical energy $ \stP{\mathcal{E}}(U)$ of a stationary point accurate to second order:
    \begin{align}
        \stP{\mathcal{E}}(U) = &\mathcal{E}(\expP{U} + \Delta U, \esP{\vec{r}} + \stP{\vec{\Delta r}}(\Delta U)) \nonumber\\
        = &{\mathcal{E}(\expP{U},\esP{\vec{r}} )}
        - \esP{q} \Delta U \nonumber\\
        + &\tfrac{1}{2} \Biggl ( \sum_{i,j,k,l} \left(\esP{\mathcal{H}_{j,k}}^{-1} \esP{\left(\tfrac{\partial q}{\partial r_k}\right)} \Delta U\right)
        \esP{\mathcal{H}_{i,j}}
        \left(\esP{\mathcal{H}_{i,l}}^{-1} \esP{\left(\tfrac{\partial q}{\partial r_l}\right)} \Delta U \right)\nonumber \\
        - &2 \sum_{i,j} \esP{\left(\tfrac{\partial q}{\partial r_i}\right)} 
        \left(\esP{\mathcal{H}_{i,j}}^{-1} \esP{\left(\tfrac{\partial q}{\partial r_j}\right)} \Delta U\right) \Delta U 
        - \esP{C_{\rm el}} \,  \Delta U^2 \Biggr) + \mathcal{O}(\Delta U^3) \quad ,\nonumber
    \end{align}
    where we dropped the force-contribution, since $\esP{\mathcal{F}_{i}} = 0$.
    Rearranging and using $\sum_i \esP{\mathcal{H}_{k,i}}^{-1}\esP{\mathcal{H}_{i,j}} = \delta_{k,j}$ 
    then yields the energy $\stP{\mathcal{E}}$ of the stationary point at potential $U = \expP{U} + \Delta U$:
    \begin{align}
        \stP{\mathcal{E}}(U) 
        = &\esP{\mathcal{E}}
        - \esP{q} \Delta U 
        - \tfrac{1}{2} \underbrace{\left(\esP{C_{\rm el}} + \overbrace{\sum_{i,j}  \esP{\mathcal{H}_{i,j}}^{-1} \esP{\left(\tfrac{\partial q}{\partial r_i}\right)} 
                \esP{\left(\tfrac{\partial q}{\partial r_j}\right)}}^{\esP{C_{\rm geom}}}\right)}_{\esP{C_{\rm total}}} \Delta U^2 + \mathcal{O}(\Delta U^3)\quad .
        \label{eq:2ndO_TE_gc_x}
    \end{align}
    This expression is essentially the (double-)integrated form of Equation~\ref{eq:total_cap_fully_analytically_gc}. We could equally obtain it by integrating Equation~\ref{eq:total_cap_fully_analytically_gc} twice from the potential $\expP{U}$ where all required properties are known (and assumed constant) to the target potential $U$.
    
    Evidently, we can also derive the excess charge in dependence of potential within this parabolic approximation:
    \begin{align}
        \stP{q}(U)
        =-\tfrac{\mathrm{d} \stP{\mathcal{E}}}{\mathrm{d}U}
        = &
        \esP{q}
        +  \left(\esP{C_{\rm el}} +\sum_{i,j}  \esP{\mathcal{H}_{i,j}}^{-1} \esP{\left(\tfrac{\partial q}{\partial r_i}\right)} 
        \esP{\left(\tfrac{\partial q}{\partial r_j}\right)}\right) \Delta U + \mathcal{O}(\Delta U^2)\quad .
        \label{eq:2ndO_TE_gc_x_q}
    \end{align}
    
    \paragraph{Discussion}
    In the following, we show that the derived expressions allow some important statements and summarize again the statements highlighted above:
    \begin{itemize}
        \item [0.]The geometric change of a stationary point in dependence of potential $U$ is well defined by the stationarity criterion (Eq.~\ref{eq:stat_criterion}) and can be considered a smooth path $\stP{\vec{r}}(U)$ for which we can state:
        \begin{itemize}
            \item [0.1] The grand canonical energy of a stationary state $\stP{\mathcal{E}}$ is given on the gcPES by $\stP{\mathcal{E}}(U)=\mathcal{E}(U, \stP{\vec{r}}(U))$.
            \item [0.2] The total derivative of $\stP{\mathcal{E}}(U)$ with respect to $U$ is given by the excess charge $q$ (Eq.~\ref{eq:dE_dU_gc_eneral}):
            \begin{align}
                \frac{\mathrm{d} \stP{\mathcal{E}}}{\mathrm{d} U}(U)= - \stP{q}(U)\quad .\nonumber
            \end{align}
            \item [0.3] The geometric shift of a stationary state is determined by the force exerted by the applied potential and the (inverse) Hessian of the system (Eq.~\ref{eq:dr_dU_general}).
            \item [0.4] The total second derivative of $\stP{\mathcal{E}}(U)$ can be derived analytically, yielding the total capacitance associated with a stationary state (Eq.~\ref{eq:total_cap_fully_analytically_gc}):
            \begin{align}
                -\tfrac{\mathrm{d}^2 \stP{\mathcal{E}}(U)}{\mathrm{d}U^2} = \stP{C_{\rm total}}(U) = \stP{C_{\rm el}}(U) + \stP{C_{\rm geom}}(U)  = \stP{C_{\rm el}} +  \stP{\nabla q}^T  \stP{\mat{\mathcal{H}}}^{-1} \stP{\nabla q}\quad.\nonumber
            \end{align}
        \end{itemize}
        \item [1.] Considering a second order expansion in potential and the geometric degrees of freedom, the grand canonical energy of any stationary point $\stP{\vec{r}}(U)$ on the gcPES has the 'classical' form of a grand canonical energy in a purely electronic second order expansion, just with an additional geometric capacitance contribution (Eq.~\ref{eq:2ndO_TE_gc_x}):
        \begin{align}     
            \stP{{\mathcal{E}}}(U) &= \esP{{\mathcal{E}}}
            - \esP{q} \Delta U 
            - \tfrac{1}{2} \left(\esP{C_{\rm el}} + \esP{C_{\rm geom}}\right) \Delta U^2  + \mathcal{O}(\Delta U^3)\qquad     \text{with}\\
            \esP{C_{\rm geom}}&=\sum_{i,j}  \esP{\mathcal{H}_{i,j}}^{-1} \esP{\left(\tfrac{\partial q}{\partial r_i}\right)} 
            \esP{\left(\tfrac{\partial q}{\partial r_j}\right)}.\label{eq:gc_geom_cap}
        \end{align}
        Similarly, the excess charge has the classical 'capacitive type' relation to the potential, however now with the additional geometric capacitance contribution (Eq.~\ref{eq:2ndO_TE_gc_x_q}):
        \begin{align}
            \stP{\Delta q}(U) =  \left(\esP{C_{\rm el}} +\esP{C_{\rm geom}}\right) \Delta U + \mathcal{O}(\Delta U^2)\quad .\nonumber
        \end{align}
        \item[2.] The geometric capacitance is separable into single contributions along the normal modes $\{n\}$ of the grand canonical Hessian $\stP{\mat{\mathcal{H}}}$:
        \begin{align}
            \stP{C_{\rm geom}} = \sum_{n} \tfrac{1}{ \stP{\mathcal{H}_{n,n}}}  \stP{\left(\tfrac{\partial q}{\partial r_n}\right)}^2\qquad ,\label{eq:geom_cap_nm_gc}
        \end{align}
        since ${\mat{\mathcal{H}}}$ becomes diagonal in its normal mode basis and so does its inverse $\mat{\mathcal{H}}^{-1}$ with ${\mathcal{H}}^{-1}_{i,j} = \tfrac{1}{{\mathcal{H}_{i,i}}} \delta_{i,j}$, reducing the double sum in ${C_{\rm geom}}$ to a single sum.
        \item[3.] Large geometric contributions originate from shallow normal modes (small $ \stP{\mathcal{H}_{n,n}}$ in Eq.~\ref{eq:geom_cap_nm_gc}) and a strong coupling between electronic and geometric degrees of freedom, i.e. large change in excess charge as a response to geometric displacement along $ \stP{\mat{\mathcal{H}}}$'s normal modes $ \stP{\left(\tfrac{\partial q}{\partial r_n}\right)}$, which is equivalent to a strong potential-dependence of the force along the normal mode $ \stP{\left(\tfrac{\partial \mathcal{F}_{n}}{\partial U}\right)}$.
        \item[4.] For a local minimum the geometric response is always positive, i.e. it leads to an increased total capacitance of the system, since at a local minimum all eigenvalues of the Hessian are positive $ \stP{\mathcal{H}_{n,n}} > 0$ and hence each element of the sum in Equation~\ref{eq:geom_cap_nm_gc} adds a positive contribution:
        \begin{align}
            \stP{C_{\rm geom, n}}&= \tfrac{1}{
                \stP{\mathcal{H}_{n,n}}
            }  \stP{\left(\tfrac{\partial q}{\partial r_{n}}\right)} ^2 > 0 \qquad .\nonumber
        \end{align}
        \item[5.] At a transition state $\ddagger$ however, the single normal mode along the reaction path $\xi$ has a negative eigenvalue (${\mathcal{H}^{\ddagger}_{\xi,\xi}} < 0$) leading to a negative contribution to the capacitance:
        \begin{align}
            \expTS{C_{\rm geom, \xi}}&= \tfrac{1}{
                \expTS{\mathcal{H}_{\xi,\xi}}
            } \expTS{\left(\tfrac{\partial q}{\partial r_{\xi}}\right)} ^2 < 0 \qquad ,\nonumber
        \end{align}
        with the perpendicular normal modes again contributing positively.
    \end{itemize}
    We want to emphasize that these statements not only have theoretical but also practical value: in principle, knowing the Hessian and the respective excess charge variations along each normal node allows to accurately derive the grand canonical energy expression of a stationary point to second order - e.g. for the analysis of thermodynamic or kinetic properties. 
    It is not even necessary to know the Hessian of the entire system, since only the normal modes with a strong coupling of geometric and electronic effects are of relevance. 
    
    \subsection{Stationary points on the canonical PES}
    While a direct grand canonical evaluation is certainly appealing, it is not always possible to perform ab-initio grand canonical calculations.
    Hence, we will evaluate the 'classical' canonical case in the following.
    This allows us a direct comparison to the grand canonical case afterwards.
    The derivation follows an identical route as before bearing in mind that $q$, $U$, and $C_{\rm el}$ have to be defined as in Table~\ref{tab:derivatives_canonical} for a consistent treatment, as will be discussed below in Section~"\nameref{sec:stat_point_gc_and_c}".
    Besides these definitions, the derivation is fully decoupled from the derivation in the grand canonical ensemble.
    \begin{table}[t]
        \centering
        \begin{tabular}{c|c|l}
            \hline
            derivative                                          &                        quantity                      &          description                   \\ \hline
            $\tfrac{\partial E}{\partial q}$                    &                 $U$                                  &         electrode potential            \\
            $\tfrac{\partial^2 E}{\partial q^2}$                &          $ \frac{1}{C_{{\rm el}}}$                   &    inverse electronic capacitance      \\
            $\tfrac{\partial E}{\partial r_i} $                 &               $- F_i$                                &            canonical forces            \\
            $\tfrac{\partial^2  E}{\partial r_j r_i}$           &                 ${H}_{ij}$                           &            canonical Hessian           \\
            $\tfrac{\partial^2 {E}}{\partial r_i \partial q}$   &            $\tfrac{\partial U}{\partial r_i}$        &   variation of electrode potential caused by spatial displacements  \\
            $=\frac{\partial^2 E}{\partial q \partial r_i}$     &     $ -\frac{\partial F_i}{\partial q} $             & variation of forces caused by changes in excess charge              \\ \hline
        \end{tabular}
        \caption{First and second derivatives of the $3N+1$ dimensional canonical energy ${E}(q, \vec{r})$.
            Keep in mind the dependence of each quantity on excess charge and geometry $(q, \vec{r})$. The mixed second derivatives are identical due to Equation~\ref{eq:grad_U_dF_dq}.}
        \label{tab:derivatives_canonical}
    \end{table}

    First, we can make the same statements using similar arguments as in the grand canonical case about the path $\stP{\vec{r}}(q)$ of a stationary point in the canonical ensemble in dependence of the electronic degree of freedom, the excess charge $q$:
    \begin{align}
        \tfrac{\mathrm{d} \stP{{E}}(q)}{\mathrm{d} q} = \tfrac{\mathrm{d}}{\mathrm{d} q} {E}(q, \stP{\vec{r}}(q))  =  \stP{\tfrac{\partial{E}}{\partial q}} + \underbrace{\sum_i  \stP{\tfrac{\partial {E}}{\partial r_i}}\cdot \tfrac{\mathrm{d} \stP{r_i}}{\mathrm{d} q}}_{=0} =  \stP{\tfrac{\partial {E}}{\partial q}} = U(q, \stP{\vec{r}}(q)) = \stP{U}(q)\quad, \label{eq:dE_dq_c_eneral}
    \end{align}
    where we used that $ \stP{\tfrac{\partial {E}}{\partial r_i}} = - \stP{F}_i=0$ for stationary points because of the (canonical) stationarity criterion:
    \begin{align}
        \stP{\vec{F}}(q) = \vec{F}(q, \stP{\vec{r}}(q)) = 0 = \rm const \qquad. \label{eq:stat_criterion_c}
    \end{align}
    We will see later, that stationary points in one ensemble are as well stationary points in the other ensemble, which allows to not differentiate between the paths of a grand canonical and canonical stationary state.
    However, note that for ease of notation, but without loss of generality, we consider $\stP{\vec{r}}$ to be a function of $q$ instead of $U$ in this section, and similarly mark other quantities using the $\stP{X}=X(q, \stP{\vec{r}}(q))$ notation.
    
    Again using the stationarity condition (Eq.~\ref{eq:stat_criterion_c}) along the entire stationary path, we can write:
    \begin{align}
        0\stackrel{!}{=} \tfrac{\mathrm{d}\stP{{F}_i}}{ \mathrm{d} q}
        = \tfrac{\mathrm{d}}{ \mathrm{d} q} {F}_i(q, \stP{\vec{r}}(q))
        &= \left( \stP{\tfrac{\partial {F}_i}{\partial q}} + \sum_j  \stP{\tfrac{\partial {F}_i}{\partial r_j}}  \tfrac{\mathrm{d} \stP{r_j}}{\mathrm{d} q}\right)
        = \left( \stP{\tfrac{\partial {F}_i}{\partial q}} - \sum_j  \stP{H_{i,j}}  \tfrac{\mathrm{d} \stP{r_j}}{\mathrm{d} q}\right) \nonumber\\
        \sum_j  \stP{H_{i,j}}  \tfrac{\mathrm{d} \stP{r_j}}{\mathrm{d} q}
        &=  \stP{\tfrac{\partial {F}_i}{\partial q}} \nonumber\\
        \frac{\mathrm{d} \stP{r_k}}{\mathrm{d} q}(q) &= \sum_i  \stP{H_{k,i}}^{-1}  \stP{\tfrac{\partial {F}_i}{\partial q}} \quad, \label{eq:dr_dq_general}
    \end{align}
    where, similar to the grand canonical case, the path is determined by the forces caused by changes in the electronic degree of freedom and the inverse (canonical) Hessian $\mat{H}$.
    Note, that we explicitly consider the force $\vec{F}$ and the Hessian $\mat{H}$ as canonical properties.
    Later, we will show that the forces are indeed identical in the two ensembles, however the Hessians might differ significantly.
    
    Now, we can derive the curvature of $\stP{E}(q)$, that we will identify as the inverse total capacitance $\stP{C_{\rm total}}(q)^{-1}$.
    That it is indeed the identical total capacitance as in the grand canonical case will become clear later.
    \begin{align}
        \stP{C_{\rm total}}(q)^{-1} &= \tfrac{\mathrm{d}^2 \stP{{E}}(q)}{\mathrm{d}q^2} 
        = \tfrac{\mathrm{d}\stP{U}(q)}{\mathrm{d} q} 
        = \tfrac{\mathrm{d}}{\mathrm{d} q}{U}(q, \stP{\vec{r}}(q)) \nonumber\\
        &=   \stP{\tfrac{\partial {U}}{\partial q}} + \sum_i  \stP{\tfrac{\partial {U}}{\partial r_i}}  \tfrac{\mathrm{d} \stP{r_i}}{\mathrm{d} q} 
        =   \stP{C_{\rm el}}^{-1} + \sum_{i,j} \stP{\tfrac{\partial {U}}{\partial r_i}} \stP{H_{i,j}}^{-1} \stP{\tfrac{\partial {F}_j}{\partial q}}\nonumber\\
        &= \stP{C_{\rm el}}^{-1} - \sum_{i,j} \stP{\tfrac{\partial {U}}{\partial r_i}} \stP{{H}_{i,j}}^{-1} \stP{\tfrac{\partial {U}}{\partial r_j}}\nonumber\\
        &= \stP{C_{\rm el}}^{-1} - \stP{\nabla U}^T \stP{\mat{{H}}}^{-1}\stP{\nabla U} \quad, \label{eq:total_cap_fully_analytically_c}
    \end{align}
    again using matrix notation and acknowledging Schwarz's theorem:
    \begin{align}
        \left.\tfrac{\partial U}{\partial r_i}\right|_q
        = \left.\tfrac{\partial }{\partial r_i} \left( \left.\tfrac{\partial E}{\partial q}\right|_r \right)\right|_q
        = \left.\tfrac{\partial }{\partial q} \left(\left.\tfrac{\partial E}{\partial r_i}\right|_q \right)\right|_r
        = - \left.\tfrac{\partial F_i}{\partial q}\right|_r \quad . \label{eq:grad_U_dF_dq}
    \end{align}
    We can again illustrate the obtained formula by considering a $(3N+1)$-dimensional parabolic expansion of the cPES $E(q, \vec{r})$.
    \paragraph{Canonical PES in the vicinity of stationary points}
    We expand the canonical energy around an arbitrary expansion point $(\expP{q}, \expP{\vec{r}})$:
    \begin{align}
        {E}(\expP{q} + \Delta q, \expP{\vec{r}} + \vec{\Delta r})
        = &\expP{E} 
        +  \expP{\left(\tfrac{\partial  E}{\partial q}\right)} \Delta q
        +  \sum_i \expP{\left(\tfrac{\partial {E}}{\partial r_i}\right)} \Delta r_i \nonumber\\
        + &\tfrac{1}{2}\left( \sum_{i,j} \Delta r_j \expP{\left(\tfrac{\partial^2 {E}}{\partial r_j \partial r_i}\right)} \Delta r_i\right. 
        + 2\sum_i\expP{\left(\tfrac{\partial^2 {E}}{\partial r_i \partial q}\right)} \Delta r_i \, \Delta q 
        +  \left. \expP{\left(\tfrac{\partial^2 {E}}{\partial q^2} \right)} \Delta q^2 \right) + \mathcal{O}(q^3, \Delta r ^3) \nonumber\\ 
        = &\expP{E}
        + \expP{U} \Delta q  - \sum_i \expP{F_{i}} \Delta r_i \nonumber\\
        + &\tfrac{1}{2}\left( \sum_{i,j} \Delta r_j \expP{{H}_{i,j}} \Delta r_i
        + 2 \sum_i \expP{\left(\tfrac{\partial U}{\partial r_i}\right)}  \Delta r_i \, \Delta q
        + \tfrac{1}{\expP{C_{\rm el}}} \,  \Delta q^2 \right) + \mathcal{O}(q^3, \Delta r ^3)  \quad , \label{eq:2ndO_TE_c}
    \end{align}
    using the definitions and relations in Table~\ref{tab:derivatives_canonical} and marking all quantities evaluated at the expansion point $(\expP{q}, \expP{\vec{r}})$ with $"\expP{ }"$. 
    
    \paragraph{Charge-induced geometric shifts of stationary points on the cPES}
    We evaluate the geometric shift of a stationary point $\Delta\stP{\vec{r}} = \stP{\vec{r}} - \esP{\vec{r}}$ caused by a change in excess charge $\Delta q = q-\expP{q}$ on the cPES expanded around a point $(\expP{q}, \esP{\vec{r}})$ is given by:
    \begin{align}
        0 \stackrel{!}{=} {F}_i(\expP{q}+\Delta q, \esP{\vec{r}} + \stP{\vec{\Delta  r}} ) 
        &=  \underbrace{\esP{F_{i}}}_{=0} + \esP{\left(\tfrac{\partial {F}_i}{\partial q}\right)} \Delta q - \sum_{j}\esP{{H}_{i,j}} \stP{\Delta r_j} + \mathcal{O}(\Delta q^2, \Delta r^2)\nonumber \\
        \Leftrightarrow 
        \stP{\Delta r_{k}} (\Delta q) &=   \sum_i \esP{{H}_{k,i}}^{-1} \esP{\left(\tfrac{\partial {F}_i}{\partial q}\right)} \Delta q + \mathcal{O}(\Delta q^2) 
        \nonumber \\
        &= - \sum_i \esP{{H}_{k,i}}^{-1} \esP{\left(\tfrac{\partial U}{\partial r_i}\right)} \Delta q + \mathcal{O}(\Delta q^2) \quad , \label{eq:delta_r_x_canonical}
    \end{align}
    where we used the interchangeability of second derivatives (Eq.~\ref{eq:grad_U_dF_dq}) in the last step.
    
    \paragraph{Canonical energy of a stationary point}
    We obtain the canonical energy of a stationary point $\stP{E}(q)$ as a function of excess charge $q$ by evaluating the expanded cPES at $q = \expP{q} + \Delta q$ and the new resting position $\esP{\vec{r}} + \stP{\vec{\Delta r}}(\Delta q)$ (Eq.~\ref{eq:delta_r_x_canonical}):
    \begin{align}
        \stP{E} (q) = &{E}(\expP{q} + \Delta q, \esP{\vec{r}} + \stP{\vec{\Delta r}}(\Delta q)) \nonumber\\
        = &{E}(\expP{q}, \esP{\vec{r}}) 
        + \esP{U} \Delta q  \nonumber\\
        + &\tfrac{1}{2}\Biggl( \sum_{i,j,n,m}
        \left(- \esP{{H}_{j,n}}^{-1} \esP{\left(\tfrac{\partial U}{\partial r_n}\right)} \Delta q\right)
        \esP{{H}_{i,j}}
        \left( - \esP{{H}_{i,m}}^{-1} \esP{\left(\tfrac{\partial U}{\partial r_m}\right)} \Delta q\right)\nonumber\\
        + &2 \sum_{i,j} \esP{\left(\tfrac{\partial U}{\partial r_i}\right)} 
        \left(- \esP{{H}_{i,j}}^{-1} \esP{\left(\tfrac{\partial U}{\partial r_j}\right)} \Delta q\right)
        \Delta q 
        + \tfrac{1}{\esP{C_{\rm el}}} \,  \Delta q^2 \Biggr) + \mathcal{O}(\Delta q^3)  \quad, 
    \end{align}
    where we again used that no forces act on stationary points ($\esP{\vec{F}}=0$). Acknowledging that $ \sum_i \esP{{H}_{k,i}}^{-1}\esP{{H}_{i,j}} = \delta_{k,j}$ and $ \esP{E} = {E}(\expP{q}, \esP{\vec{r}})$:
    \begin{align}
        \stP{E}(q)
        = \esP{E}
        + \esP{U} \Delta q  
        + \tfrac{1}{2}\left(\tfrac{1}{\esP{C_{\rm el}}} - \sum_{i,j}  \esP{H_{i,j}}^{-1} \esP{\left(\tfrac{\partial U}{\partial r_i}\right)} 
        \esP{\left(\tfrac{\partial U}{\partial r_j}\right)} 
        \right) \,  \Delta q^2 + \mathcal{O}(\Delta q^3) \quad ,  \label{eq:2ndO_TE_c_x}
    \end{align}
    with its first derivative, the electrode potential $U$:
    \begin{align}
        \stP{U}(q)
        = \tfrac{\mathrm{d} \stP{E}}{\mathrm{d} q}
        = \esP{U}
        + \left(\tfrac{1}{\esP{C_{\rm el}}} - \sum_{i,j}  \esP{H_{i,j}}^{-1} \esP{\left(\tfrac{\partial U}{\partial r_i}\right)} 
        \esP{\left(\tfrac{\partial U}{\partial r_j}\right)} 
        \right) \,  \Delta q + \mathcal{O}(\Delta q^2) \quad . \label{eq:2ndO_TE_c_x_q}
    \end{align}

    \paragraph{Discussion}\label{sec:c_discussion}
    Before we discuss the slightly more complex expression for the quadratic prefactor in the energy expression, we first summarize the general statements for the canonical PES, similar as in the grand canonical case, given above:
    \begin{itemize}
        \item [0.] A stationary state in the cPES is well defined by the stationarity criterion (Eq.~\ref{eq:stat_criterion_c}) and its path in dependence of excess charge can be considered as a smooth path $\stP{\vec{r}}(q)$.
        \begin{itemize}
            \item [0.1] The canonical energy of a stationary state $\stP{E}(q)$ is defined on the cPES via $\stP{E}(q) = E(q, \stP{\vec{r}}(q))$.
            \item [0.2] Its total derivative with respect to $q$ is given by the electrode potential $U$ (Eq.~\ref{eq:dE_dq_c_eneral}):
            \begin{align}
                \frac{\mathrm{d} \stP{{E}}}{\mathrm{d} q}(q) = \stP{U}(q) \quad .\nonumber
            \end{align}
            \item [0.3] The geometric change induced by a change in excess charge originates from the caused forces and is affected by the inverse (canonical) Hessian (Eq.~\ref{eq:dr_dq_general}).
            \item [0.4] The (canonical) curvature of the canonical energy along the stationary path  on the cPES, that we will denote as ${\gamma_{\rm total}}$, is given by:
            \begin{align}
                \tfrac{\mathrm{d}^2 \stP{{E}}}{\mathrm{d}q^2} = \stP{\gamma_{\rm total}}  
                =\stP{C_{\rm el}}^{-1} - \stP{\nabla U}^T \stP{\mat{{H}}}^{-1}\stP{\nabla U}\quad ,\nonumber
            \end{align}
            where again every quantity has to be considered a function of $q$.
        \end{itemize}
    \end{itemize}
    The prefactor of the quadratic term, the curvature $\stP{\gamma_{\rm total}}$, has a similar form as in the grand canonical case, where it defines the negative total capacitance of the system ${\tfrac{\mathrm{d}^2 \stP{\mathcal{E}}}{\mathrm{d} U^2}} = -\stP{C_{\rm total}}$.
    However, the situation is slightly more complex in the canonical case, since, here, capacitances enter inversely, similar as e.g. the purely electronic capacitance is the inverse of the purely electronic curvature $\tfrac{\partial^2 \stP{E}}{\partial q^2} = C_{\rm el}^{-1}$.
    We follow this thought and identify the total curvature as the inverse of the total capacitance:
    \begin{align}
        \tfrac{1}{\stP{C_{\rm total}}} &= \tfrac{1}{\stP{C_{\rm el}}} - \sum_{i,j}  \stP{H_{i,j}}^{-1} \stP{\left(\tfrac{\partial U}{\partial r_i}\right)} 
        \stP{\left(\tfrac{\partial U}{\partial r_j}\right)} \\
        \Leftrightarrow \stP{C_{\rm total}} &= \stP{C_{\rm el}}\, \frac{1}{1 - \stP{C_{\rm el}}\sum_{i,j}  \stP{{H}_{i,j}}^{-1} \stP{\left(\tfrac{\partial U}{\partial r_i}\right)} 
            \stP{\left(\tfrac{\partial U}{\partial r_j}\right)}}\quad . \label{eq:C_total_from_c_as_frac}
    \end{align}
    And indeed, we will show later that this identification is right, i.e. consistent with the grand canonical case.
    Note, that the inverse sum forming the total capacitance is the reason why we can not easily identify $\stP{C_{\rm geom}}^{-1} = \sum_{i,j}  \stP{H_{i,j}}^{-1} \stP{\left(\tfrac{\partial U}{\partial r_i}\right)} \stP{\left(\tfrac{\partial U}{\partial r_j}\right)}$. This will become clearer in the discussion of the relation between canonical and grand canonical results below.
    
    We first discuss this expression in terms of the (canonical) curvatures $\stP{\gamma_{\rm total}} = \stP{C_{\rm total}}^{-1}$,  $\quad \stP{\gamma_{\rm el}} =  \stP{C_{\rm el}}^{-1}$, and $\stP{\gamma_{\rm geom}} =  \sum_{i,j}  \stP{{H}_{i,j}}^{-1} \stP{\left(\tfrac{\partial U}{\partial r_i}\right)} \stP{\left(\tfrac{\partial U}{\partial r_j}\right)}$, yielding the following relations:
    \begin{align}
        \stP{\gamma_{\rm total}} &= \stP{\gamma_{\rm el}} - \stP{\gamma_{\rm geom}}
        = \stP{\gamma_{\rm el}}  \left(1-\tfrac{\stP{\gamma_{\rm geom}}}{\stP{\gamma_{\rm el}}}\right)\label{eq:tot_curv_c}\\
        \stP{C_{\rm total}}
        &= \stP{C_{\rm el}}  \left(1-\tfrac{\stP{\gamma_{\rm geom}}}{\stP{\gamma_{\rm el}}}\right)^{-1} \nonumber\quad.
    \end{align}
    We can think of the ratio $\tfrac{\stP{\gamma_{\rm geom}}}{\stP{\gamma_{\rm el}}}$ as a measure of the strength of the geometric effects relative to the electronic ones.
    This can again be seen by considering the normal modes of the (canonical) Hessian,
    \begin{align}
        \tfrac{\stP{\gamma_{\rm geom}}}{\stP{\gamma_{\rm el}}} = \stP{C_{\rm el}}\sum_{n}  \tfrac{1}{\stP{{H}_{n,n}}} \stP{\left(\tfrac{\partial U}{\partial r_n}\right)}^2    \quad,\label{eq:geom_curv_c}
    \end{align}
    in which the double sum becomes a single sum, and again is separable into contributions along the (canonical) normal modes.
    While the Hessian's eigenvalues $\stP{H_{n,n}}$ can be thought of as the curvature in the 'chemical' energy when displacing the system along a normal mode, we can think of $\stP{C_{\rm el}}\stP{\left(\tfrac{\partial U}{\partial r_n}\right)}^2$ as the curvature of the 'electrostatic energy' originating from the displacement.
    We can make the following statements based on the derived canonical-based expression for the total capacitance (for these statements we assume a 'normal' electronic response, i.e. $\stP{C_{\rm el}} > 0$):
    \begin{itemize}
        \item[1.] Similar to the grand canonical case, large geometric contributions $\stP{\gamma_{\rm geom}}$ originate from shallow normal modes (small $\stP{{H}_{n,n}}$) and large changes in electrode potential as a response to geometric displacement $\stP{\left(\tfrac{\partial U}{\partial r_n}\right)}$ or equivalently a strong potential-dependence of the force along the normal mode $\stP{\left(\tfrac{\partial {F}_{n}}{\partial q}\right)}$ (see  Eqs.~\ref{eq:geom_curv_c} and~\ref{eq:grad_U_dF_dq}).
        \item[2.] Local minima always exhibit a positive geometric curvature ($\stP{\gamma_{\rm geom}}>0$), since the eigenvalues of the Hessian and, hence, every contribution of any normal mode is positive (see Eq.~\ref{eq:geom_curv_c}). This decreases the total curvature and might cause it to switch its sign and become negative (see Eq.~\ref{eq:tot_curv_c}).
        \item[3.] At a transition state $\ddagger$, the single normal mode along the reaction path $\xi$ has a negative eigenvalue ($\stP{{H}_{\xi,\xi}}^{\ddagger} < 0$) leading to a negative geometric contribution to $\stP{\gamma_{\rm geom}}$ and an increasing contribution to the total curvature (Eqs.~\ref{eq:geom_curv_c} and~\ref{eq:tot_curv_c}).
        \item[4.] Depending on the relative strength of geometric and electronic effects, i.e. the ratio between geometric curvature $\stP{\gamma_{\rm geom}}$ and electronic curvature $\stP{\gamma_{\rm el}}$, we have to consider two cases:\\
        Dominating electronic effects: $\stP{\gamma_{\rm el}}>\stP{\gamma_{\rm geom}}$:
        \begin{itemize}
            \item [4.1] The total curvature $\stP{\gamma_{\rm total}}$ is always positive and and decreases with increasing geometric effects.
            \item [4.2] The total capacitance is, hence, always positive and increases with increasing geometric effects. This can also be seen by expanding $\stP{C_{\rm total}} = \stP{C_{\rm el}}\left(1- \tfrac{\stP{\gamma_{\rm geom}}}{\stP{\gamma_{\rm el}}}\right)^{-1}$ in its Taylor series:
            \begin{align}
                \stP{C_{\rm total}} &= \stP{C_{\rm el}}\left(1+\tfrac{\stP{\gamma_{\rm geom}}}{\stP{\gamma_{\rm el}}}+\left(\tfrac{\stP{\gamma_{\rm geom}}}{\stP{\gamma_{\rm el}}}\right)^2+...\right)\nonumber\\
                &\approx \stP{C_{\rm el}} + \stP{C_{\rm el}}^2\sum_{i,j}  \stP{{H}_{i,j}}^{-1} \stP{\left(\tfrac{\partial U}{\partial r_i}\right)} \stP{\left(\tfrac{\partial U}{\partial r_j}\right)} + ... \quad .\nonumber
            \end{align}
            \item[4.3] For local minima geometric effects lead to an increased total capacitance -- identical to the grand canonical case.
        \end{itemize}
        Dominating geometric effects $\stP{\gamma_{\rm geom}}>\stP{\gamma_{\rm el}}$:
        \begin{itemize}
            \item[4.4] The total curvature $\stP{\gamma_{\rm total}}$ is always negative and decreases further with increasing geometric effects.
            \item[4.5] The total capacitance is always negative and increases (becomes less negative) with increasing geometric effects.
            \item[4.6] Local minima exhibit a negative total capacitance -- in contrast to the grand canonical case. While intuitively an increase in excess charge should lead to an increase in potential (which is the case for a positive capacitance), the additional geometric contribution leads here to a decrease in potential with increasing excess charge (which is reflected in a negative capacitance).
        \end{itemize}
    \end{itemize}
    Now, one might legitimately wonder, why a local minimum in the canonical ensemble can exhibit a negative total capacitance in case of strong geometric effects, while this is not possible for a local minimum in the grand canonical case. We will resolve this apparent contradiction in the following by relating the results in both ensembles. There, we also discuss the different cases depending on the relative strength of geometric and electronic effects in more detail.

    \subsection{Relation between stationary points in the canonical and the grand canonical ensemble}\label{sec:stat_point_gc_and_c}
    In order to analyze the relationship between stationary points in both the grand canonical and canonical ensemble, we start with important general relations between canonical and grand canonical quantities.
    \paragraph{Grand canonical and canonical energy}
    The grand canonical energy $\mathcal{E}(U, \vec{r})$ and the canonical energy $E(q,\vec{r})$ are related via a Legendre-transformation in the conjugate variables $U$ and $q$ respectively (see e.g. Refs.~\citenum{kastlunger2018controlled, hormann2019grand, hormann2020electrosorption, ringe2022implicit}):
    \begin{align}
        \mathcal{E}(U,\vec{r}) = E(q(U, \vec{r}), \vec{r}) - q(U, \vec{r})U
        \label{eq:LT_c_to_gc_generic}\quad,
    \end{align}
    where $U$ is defined as $U = \left.\tfrac{\partial E}{\partial q}\right|_r = U(q, \vec{r})$ and $q$ as its inverse function $q(U, \vec{r})$. Note that from these definitions, it directly follows, that:
    \begin{align}
        \left.\tfrac{\partial \mathcal{E}}{\partial U}\right|_r =  \left.\tfrac{\partial}{\partial U}\right|_r  \left(E(q(U, \vec{r}), \vec{r}) - q(U, \vec{r})U)\right) = \left.\tfrac{\partial E}{\partial q}\right|_r \cdot \left.\tfrac{\partial q}{\partial U}\right|_r     - \left.\tfrac{\partial q}{\partial U}\right|_r U - q  =  -q\quad .\nonumber
    \end{align}
    This justifies using the same quantities $q$ and $U$ in the two ensembles (i.e. in Tables \ref{tab:derivatives} and \ref{tab:derivatives_canonical}).
    Note, however, that in the canonical ensemble $q$ is the independent variable and $U$ the dependent variable being a function of $q$: \,  $U=U(q,\vec{r})$ and vice versa in the grand canonical case.
    Furthermore it is essential, that defining the purely electronic capacitance in the the grand canonical ensemble as $C_{\rm el} = -\tfrac{\partial^2 \mathcal{E}}{\partial U^2}$ fixes the canonical one as $\tfrac{\partial^2 E}{\partial q^2} = C_{\rm el}^{-1}$, by construction of the Legendre transform:
    \begin{align}
        1 = \left.\tfrac{\partial q(U(q, \vec{r}), \vec{r})}{\partial q}\right|_r =\left. \tfrac{\partial q}{\partial U}\right|_r \cdot \left.\tfrac{\partial U}{\partial q}\right|_r =  - \left.\tfrac{\partial^2 \mathcal{E}}{\partial U^2}\right|_r \cdot \left.\tfrac{\partial^2 E}{\partial q^2}\right|_r = C_{\rm el} \left. \tfrac{\partial^2 E}{\partial q^2}\right|_r\quad.\nonumber
    \end{align}
    This justifies that we use the same electronic capacitance in both ensembles (see Tables \ref{tab:derivatives} and \ref{tab:derivatives_canonical}).
    
    \paragraph{Grand canonical and canonical Forces}
    Similarly, we can directly evaluate the relation between grand canonical and canonical Forces ($\vec{\mathcal{F}}$ and $\vec{F}$) defined as the negative spatial gradient of the respective energies.
    The $i$-th component of the grand canonical Force $\vec{\mathcal{F}}(U,\vec{r})$ is given by:
    \begin{align}
        \mathcal{F}_i(U,\vec{r}) & = - \left.\tfrac{\partial \mathcal{E}}{\partial r_i}\right|_{U} 
        =  - \left.\tfrac{\partial}{\partial r_i} \biggl (E(q(U, \vec{r}), \vec{r}) - q(U,\vec{r})U\biggr)\right|_{U} \nonumber\\
        & = - \biggl (\left.\tfrac{\partial E}{\partial r_i}\right|_{q}  + \underbrace{\left.\tfrac{\partial E}{\partial q}\right|_{r}}_{=U} \cdot \left.\tfrac{\partial q}{\partial r_i}\right|_{U}  -  \left.\tfrac{\partial q}{\partial r_i}\right|_{U}U) \biggr)
        = - \left.\tfrac{\partial E}{\partial r_i}\right|_{q} 
        = F_i(q(U, \vec{r}), \vec{r})\quad. \label{eq:trafo_forces}
    \end{align}
    Hence, grand canonical and canonical forces are identical when evaluated at the respective independent variables.
    This was also numerically analyzed in detail in Reference~\citenum{lindgren2022electrochemistry}.
    Again, note that $\mathcal{E}$, $\mathcal{F}$, and $q$ are functions of $U$ and $\vec{r}$ and that $E$, $F$, and $U$ are functions of $q$ and $\vec{r}$.
    
    In the following, for ease of notation we do not mark the dependencies any more.
    However keep in mind, that in equations like above, if the left side is a function of $(U, \vec{r})$ but the right side is one of $(q, \vec{r})$, then we have to consider the latter a function of $(q(U, \vec{r}), \vec{r})$, if we want to be rigorous.
    
    \paragraph{Electronic and Geometric Coupling Terms}
    In order to derive the relations between the mixed derivatives in $U$, $q$, and $\vec{r}$ we can make use of the triple product acknowledging that both quantities $U=U(q, \vec{r})$ and $q=q(U, \vec{r})$ can be identically expressed via each other:
    \begin{align}
        -1 =  \left.\tfrac{\partial r_i}{\partial q}\right|_U  \left.\tfrac{\partial q}{\partial U}\right|_r \left.\tfrac{\partial U}{\partial r_i}\right|_q  &\Leftrightarrow
        \left.\tfrac{\partial q}{\partial r_i}\right|_U 
        = - C_{\rm el} \left.\tfrac{\partial U}{\partial r_i}\right|_q \qquad \text{ hence:} \label{eq:triple_product}\\
        \left.\tfrac{\partial q}{\partial r_i}\right|_U   
        =  \left.\tfrac{\partial \mathcal{F}_i}{\partial U}\right|_r 
        &= - C_{\rm el} \left.\tfrac{\partial U}{\partial r_i}\right|_q  
        =  C_{\rm el} \left.\tfrac{\partial F_i}{\partial q}\right|_r \quad, \nonumber
    \end{align}
    remembering that grand canonical quantities $q$ and $\vec{\mathcal{F}}$ and their derivatives are functions of $(U, \vec{r})$ and similar the canonical ones $U$ and $\vec{F}$ and their derivatives functions of $(q, \vec{r})$ and using the above discussed equivalence of the respective mixed second derivatives (Eqs.~\ref{eq:grad_q_dF_dU} and~\ref{eq:grad_U_dF_dq}).
    Note, that the derivation using the triple product could instead similarly be approached analogous to that of the Forces (Eq.~\ref{eq:trafo_forces}).
    
    \paragraph{Grand canonical and canonical Hessians}\label{sec:gc-c-Hessians}
    The relation between the grand canonical Hessian $\mathcal{H}_{i,j}$ and the canonical Hessian $H_{i,j}$ will play a central role in discussing the apparent contradiction, that the geometric response in the canonical ensemble can lead to negative total capacitances, while this is not possible in the grand canonical case. We can relate the Hessians by:
    \begin{align}
        \mathcal{H}_{i,j} & = \left.\tfrac{\partial^2 \mathcal{E}}{\partial r_j \partial r_i}\right|_U 
        = \left.\tfrac{\partial }{\partial r_j} \left( \left.\tfrac{\partial \mathcal{E}}{\partial r_i}\right|_U \right)\right|_U \nonumber \\
        & = \left.\tfrac{\partial }{\partial r_j} \left( \left.\tfrac{\partial E}{\partial r_i}\right|_q \right)\right|_U 
        = \left.\tfrac{\partial }{\partial r_j} \left( \left.\tfrac{\partial E}{\partial r_i}\right|_q \right)\right|_q 
        +  \left.\tfrac{\partial }{\partial q} \left( \left.\tfrac{\partial E}{\partial r_i}\right|_q \right)\right|_r \cdot \left.\tfrac{\partial q}{\partial r_j}\right|_U  \nonumber\\
        &= \left.\tfrac{\partial^2 E}{\partial r_j\partial r_i}\right|_q 
        + \left.\tfrac{\partial }{\partial r_i} \left( \left.\tfrac{\partial E}{\partial q}\right|_r \right)\right|_q  \cdot \left.\tfrac{\partial q}{\partial r_j}\right|_U  
        = H_{i,j} +  \left.\tfrac{\partial U}{\partial r_i}\right|_q \cdot \left.\tfrac{\partial q}{\partial r_j}\right|_U \quad,
    \end{align}
    using the equivalence of forces (Eq.~\ref{eq:trafo_forces}),  the canonical definition of the electrode potential $\left.\tfrac{\partial E}{\partial q}\right|_r = U$, Schwarz's theorem $\left.\tfrac{\partial }{\partial r_i} \left( \left.\tfrac{\partial E}{\partial q}\right|_r \right)\right|_q = \left.\tfrac{\partial }{\partial q} \left( \left.\tfrac{\partial E}{\partial r_i}\right|_q \right)\right|_r$,  and again considering that $\mathcal{E}$ and its derivatives, i.e. $\mat{\mathcal{H}}$ and $q$, are functions of $(U, \vec{r})$ and that $E$ and its derivatives, i.e. $\mat{H}$ and $U$, are functions of $(q, \vec{r})$.
    
    Note, that we can rewrite $\left.\tfrac{\partial q}{\partial r_j}\right|_U$ using the just derived triple product (Eq.~\ref{eq:triple_product}) leading to an expression that is purely canonical, i.e. only depending on $(q, \vec{r})$:
    \begin{align}
        \mathcal{H}_{i,j} = H_{i,j} - C_{\rm el} \cdot \left.\tfrac{\partial U}{\partial r_i}\right|_q \cdot \left.\tfrac{\partial U}{\partial r_j}\right|_q \label{eq:gc-Hessian-from-c}\\
        \mat{\mathcal{H}} = \mat{H} -  C_{\rm el}\nabla U \nabla U^T\quad,\nonumber
    \end{align}
    where we also show the corresponding matrix notation. Note the $\nabla U \nabla U^T$ denotes an outer product, and that $\mat{\mathcal{H}}$ on the left side has to be considered a function of $(U,\vec{r})$ and all properties on the right a function of $(q(U, \vec{r}), \vec{r})$.
    
    This has the important consequence, that if we calculate the canonical Hessian $\mat{H}$ at constant charge conditions $({q}, {\vec{r}})$ using a finite difference approach displacing each atom in each spatial coordinate, and also get the potential (work function) changes caused by the displacements $\nabla U$, we can compute the grand canonical Hessian $\mat{\mathcal{H}}$ directly as long as the capacitance $C_{\rm el}$ is known.
    
    \paragraph{Inverse relationships of Hessians}
    The Hessians enter the capacitance expressions (Eqs.~\ref{eq:total_cap_fully_analytically_gc} and~\ref{eq:total_cap_fully_analytically_c}) in an inverse manner.
    This poses a challenge since there are no general analytical expressions for the inverse of a sum of two matrices.
    However, for the discussed case an analytical expression does exist.
    This is due to the nature of the second term in Equation~\ref{eq:gc-Hessian-from-c} that we rewrote as $-C_{\rm el}\nabla U \nabla U^T$, i.e. as outer product of two (identical) vectors, hence the matrix defined by this expression has rank one.
    Now, for the inverse of a sum of two matrices $\mat{A}$ and $\mat{B}$, where $\mat{A}$ and $\mat{A}+\mat{B}$ are invertible (which is in general the case for the canonical and the grand canonical Hessian) and where $\mat{A}$ has rank one, then $\mat{A}+\mat{B}$ is given by (see e.g. Ref.~\citenum{miller1981inverse}):
    \begin{align}
        (\mat{A}+\mat{B})^{-1} =\mat{A}^{-1} - \frac{1}{1+\mathrm{tr}(\mat{B}\,\mat{A}^{-1})}\mat{A}^{-1}\mat{B}\,\mat{A}^{-1}\quad,\nonumber
    \end{align}
    which yields in our case:
    \begin{align}
        \mathcal{H}^{-1}_{i,j} &= H^{-1}_{i,j} + \frac{C_{\rm el}}{1 - C_{\rm el}\sum_{k,l} H^{-1}_{k,l} \tfrac{\partial U}{\partial r_k}\tfrac{\partial U}{\partial r_l}}  \sum_{n,m}  H^{-1}_{i,n}  \tfrac{\partial U}{\partial r_n} \tfrac{\partial U}{\partial r_m} H^{-1}_{m,j}\label{eq:inv_hessians_gc_and_c}\\
        \mat{\mathcal{H}}^{-1} &= \mat{H}^{-1} + \tfrac{C_{\rm el}}{1-C_{\rm el} \nabla U ^T \mat{{H}}^{-1} \nabla U}  \mat{{H}}^{-1} \nabla U \nabla U ^T  \mat{{H}}^{-1} \quad.\nonumber
    \end{align}
    While the use of this rather complicated formula might not directly be apparent, it will play an essential role in expressing the geometric response, i.e. the geometric contribution to the capacitance, in terms of canonical properties.
    
    \paragraph{Relation between stationary points and their geometric responses}
    Since forces are identical in both ensembles (Eq.~\ref{eq:trafo_forces}), a point $(\esP{U}, \esP{\vec{r}})$ in the grand canonical ensemble is stationary if and only if it its counterpart $(\esP{q}, \esP{\vec{r}}) = (q(\esP{U}, \esP{\vec{r}}), \esP{\vec{r}})$ is stationary in the canonical ensemble (as already discussed by others in Refs.~\citenum{lindgren2022electrochemistry, bureau1997modeling}:
    \begin{align}
        \esP{\vec{\mathcal{F}}} = \vec{\mathcal{F}} (\esP{U}, \esP{\vec{r}}) = 0 \qquad 
        \Leftrightarrow \qquad \esP{\vec{F}} = \vec{F} (\esP{q}, \esP{\vec{r}}) = 0 \quad. \label{eq:stat_point_in_both_ensembles}
    \end{align}
    This justifies that we do not differentiate between canonical and grand canonical stationary points, i.e. why we use the same notation $\stP{\vec{r}}$.
    The canonical and the grand canonical paths are identical -- it does not matter whether we calculate $\stP{\vec{r}}$ at canonical, constant charge conditions (obtaining $\stP{U}(q)$ as an output) or at grand canonical constant potential conditions (obtaining $\stP{q}(U)$ as an output). 
    
    However, even though this states that a stationary point stays a stationary point while switching the ensemble, it does not yet make any statements about the 'character' of the stationary point, i.e. whether a local minimum in one ensemble is also always a local minimum in the other one or maybe even becomes a transition state.
    In order to evaluate this, let us Legendre-transform the second order expansion around a point in the cPES (Eq.~\ref{eq:2ndO_TE_c}) to the expansion of the gcPES (Eq.~\ref{eq:2ndO_TE_gc}) (the reverse would also be possible).
    This yields the identical expression as when directly expanded in the grand canonical ensemble.
    The canonical and the grand canonical expansion essentially carry the same information.
    While this might appear mathematically reasonable without any further proof, we want to show this explicitly, by
    \begin{align}
        E(q, \vec{r}) &= E(\expP{q} + \Delta q, \expP{\vec{r}} + \vec{\Delta r}) \nonumber \\
        = &\expP{E} 
        + \expP{U} \Delta q  - \sum_i \expP{F_{i}} \Delta r_i 
        + \tfrac{1}{2}\left( \sum_{i,j} \Delta r_j \expP{{H}_{i,j}} \Delta r_i
        + 2 \sum_i \expP{\left(\tfrac{\partial U}{\partial r_i}\right)}  \Delta r_i \, \Delta q
        + \tfrac{1}{\expP{C_{\rm el}}} \,  \Delta q^2 \right) \nonumber\\
        + &\mathcal{O}(\Delta q^3, \Delta r^3)\nonumber \\
        \Rightarrow    U &= \tfrac{\partial E}{\partial q}  
        = \expP{U} + \tfrac{1}{\expP{C_{\rm el}}}(q-\expP{q}) + \sum_i \expP{\left(\tfrac{\partial U}{\partial r_i}\right)} \Delta r_i
        + \mathcal{O}(\Delta q^2, \Delta r^2) \nonumber \\
        \Leftrightarrow  q &= \expP{q} + \expP{C_{\rm el}} (U-\expP{U} 
        - \sum_i \expP{\left(\tfrac{\partial U}{\partial r_i}\right)} \Delta r_i) 
        + \mathcal{O}(\Delta U^2, \Delta r^2)\nonumber\\ 
        \Rightarrow   \mathcal{E}(U, \vec{r}) & = E(q(U, \vec{r}), \vec{r}) - q(U, \vec{r})U\nonumber\\
        & = \mathcal{E}_0 -\expP{q}\Delta U - \sum_i \expP{F_{i}} \Delta r_i\nonumber\\
        & + \tfrac{1}{2} \left( \sum_{i,j} \Delta r_j \left[\expP{H_{i,j}} - \expP{C_{\rm el}} \expP{\left(\tfrac{\partial U}{\partial r_i}\right)}\expP{\left(\tfrac{\partial U}{\partial r_j}\right)}\right]\Delta r_i
        + 2 \sum_i \expP{C_{\rm el}}  \expP{\left(\tfrac{\partial U}{\partial r_i}\right)} \Delta r_i \Delta U
        - \expP{C_{\rm el}} \Delta U^2 \right)\nonumber \\
        & + \mathcal{O}(\Delta U^3, \Delta r^3)\quad, \nonumber
    \end{align}
    using the Legendre transformation (Eq.~\ref{eq:LT_c_to_gc_generic}) and $\expP{\mathcal{E}} = \expP{E} - \expP{q} \expP{U}$ and performing some rearrangements.
    Considering the equality of forces (Eq.~\ref{eq:trafo_forces}), the triple product (Eq.~\ref{eq:triple_product}), and the relation between canonical and grand canonical Hessians (Eq.~\ref{eq:gc-Hessian-from-c}), this is identical to the second order expression purely derived in the grand canonical ensemble (Eq.~\ref{eq:2ndO_TE_gc}).
    
    This has the important consequence, that any quantity derived above in the gcPES can equally be derived in the cPES and vice versa.
    Hence, deriving the grand canonical energy of a stationary point $\stP{\mathcal{E}}(U)$ (Eq.~\ref{eq:2ndO_TE_gc_x}) from the canonical case, i.e. from the canonical energy of a stationary point $\stP{E}(q)$ (Eq.~\ref{eq:2ndO_TE_c_x}) has to yield identical results.
    Indeed, we can follow this route and Legendre-transform $\stP{E}(q)$ to the grand canonical ensemble:
    \begin{align}
        \stP{E}(q)
        &= \esP{E}
        + \esP{U} \Delta q  
        + \tfrac{1}{2}\left(\tfrac{1}{\esP{C_{\rm el}}} - \sum_{i,j}  \esP{H_{i,j}}^{-1} \esP{\left(\tfrac{\partial U}{\partial r_i}\right)} 
        \esP{\left(\tfrac{\partial U}{\partial r_j}\right)} 
        \right) \,  \Delta q^2 + \mathcal{O}(\Delta q^3) \nonumber\\
        \Rightarrow U &= \tfrac{\partial \stP{E}}{\partial q} = \esP{U} + \left(\tfrac{1}{\esP{C_{\rm el}}} - \sum_{i,j}  \esP{H_{i,j}}^{-1} \esP{\left(\tfrac{\partial U}{\partial r_i}\right)} 
        \esP{\left(\tfrac{\partial U}{\partial r_j}\right)} 
        \right)\left(q-\expP{q}\right) + \mathcal{O}(\Delta q^2)\nonumber\\
        \Leftrightarrow  q &= \expP{q} +  \left(\tfrac{1}{\esP{C_{\rm el}}} - \sum_{i,j}  \esP{H_{i,j}}^{-1} \esP{\left(\tfrac{\partial U}{\partial r_i}\right)} 
        \esP{\left(\tfrac{\partial U}{\partial r_j}\right)} 
        \right)^{-1} \left(U-\expP{U}\right) + \mathcal{O}(\Delta U^2)\nonumber\\
        \Rightarrow \stP{\mathcal{E}}(U) &=     \esP{\mathcal{E}} - \esP{q}\Delta U - \tfrac{1}{2} \underbrace{\left(\tfrac{1}{\esP{C_{\rm el}}} - \sum_{i,j}  \esP{H_{i,j}}^{-1} \esP{\left(\tfrac{\partial U}{\partial r_i}\right)} 
            \esP{\left(\tfrac{\partial U}{\partial r_j}\right)} 
            \right)^{-1}}_{=\esP{C_{\rm total}}}  \Delta U ^2 + \mathcal{O}(\Delta U^3)\label{eq:gc_x_from_c_x_LT}\quad,
    \end{align}
    using the Legendre transformation (Eq.~\ref{eq:LT_c_to_gc_generic}) and $\esP{\mathcal{E}} = \esP{E} - \esP{U}\esP{q}$.
    This yields the identical result as in the purely grand canonical case if we relate the total capacitance to the negative curvature. Generalized this means:
    \begin{align}
        \stP{C_{\rm total}} &= \left(\tfrac{1}{\stP{C_{\rm el}}} - \sum_{i,j}  \stP{{H}_{i,j}}^{-1} \stP{\left(\tfrac{\partial U}{\partial r_i}\right)} 
        \stP{\left(\tfrac{\partial U}{\partial r_j}\right)}
        \right)^{-1} \quad \text{(from cPES)}\label{eq:discussion_gc_geom_cap_first_line}\\
        &=\stP{C_{\rm el}} + \sum_{i,j}  \stP{\mathcal{H}_{i,j}}^{-1} \stP{\left(\tfrac{\partial q}{\partial r_i}\right)} 
        \stP{\left(\tfrac{\partial q}{\partial r_j}\right)} \qquad \text{(from gcPES, Eq.~\ref{eq:gc_geom_cap})}. \label{eq:discussion_gc_geom_cap}
    \end{align}
    The different mathematical forms when expressed in canonical or grand canonical quantities and their intricate relationships originate from three facts:
    \begin{itemize}
        \item [1.] Geometric effects enter as additive terms in the curvatures, which is in essence the total capacitance in the grand canonical ensemble but the inverse of the total capacitance in the canonical ensemble.
        \item [2.] Only the inverses $\mat{H}^{-1}$ and $\mat{\mathcal{H}}^{-1}$ of the respective Hessian enter in the geometric contributions (Eqs.~\ref{eq:total_cap_fully_analytically_gc} and~\ref{eq:total_cap_fully_analytically_c}).
        \item [3.] The grand canonical Hessian differs from the canonical one by an additive term (see Eq.~\ref{eq:gc-Hessian-from-c}).
    \end{itemize} 
    Indeed, using the relation between the Hessians (Eq.~\ref{eq:gc-Hessian-from-c}), it can be shown that both expressions for the total capacitance (Eqs.~\ref{eq:discussion_gc_geom_cap_first_line} and \ref{eq:discussion_gc_geom_cap}) are identical.
    The interested readers might convince themselves of this, by assuming a one-dimensional problem where the Hessians reduce to scalars and hence the two expressions can be easily converted into each other (i.e. by showing that $\left(\tfrac{1}{c} - \tfrac{\dot{U}^2}{(\mathcal{H} + c\dot{U}^2)}\right)^{-1} = c+ \tfrac{c^2\dot{U}^2}{\mathcal{H}}$).
    In the high-dimensional case, with the Hessians being $3N\times3N$ matrices, showing the equivalence is a bit more elaborate, but possible using the above derived relation between the inverses of the Hessians (Eq.~\ref{eq:inv_hessians_gc_and_c}) and the triple product (Eq.~\ref{eq:triple_product}).
    Starting from the grand canonical expression (Eq.~\ref{eq:discussion_gc_geom_cap}) and using matrix notation:
    \begin{align}
        \stP{C_{\rm total}} &=  \stP{C_{\rm el}} + \stP{\nabla q}^T \stP{\mat{\mathcal{H}}}^{-1} \stP{\nabla q} \nonumber\\
        &=  \stP{C_{\rm el}} + \stP{C_{\rm el}}^2 \stP{\nabla U} ^T \stP{\mat{\mathcal{H}}}^{-1} \stP{\nabla U} \nonumber\\
        &=  \stP{C_{\rm el}} +  \stP{C_{\rm el}}^2 \stP{\nabla U} ^T \left(\stP{\mat{H}}^{-1} + \tfrac{ \stP{C_{\rm el}}}{1- \stP{C_{\rm el}}\stP{\nabla U} \stP{\mat{H}}^{-1}\stP{\nabla U}} \stP{\mat{H}}^{-1}\stP{\nabla U} \stP{\nabla U}^T\stP{\mat{H}}^{-1}\right)\stP{\nabla U} \nonumber\\
        &=  \stP{C_{\rm el}} +  \stP{C_{\rm el}}\left(\stP{C_{\rm el}}\stP{\nabla U}^T \stP{\mat{H}}^{-1}\stP{\nabla U} + \tfrac{\left(\stP{C_{\rm el}}\stP{\nabla U}^T \stP{\mat{H}}^{-1}\stP{\nabla U}\right)^2}{1- \stP{C_{\rm el}}\stP{\nabla U}^T \stP{\mat{H}}^{-1}\stP{\nabla U}}\right)\nonumber\\
        &=  \stP{C_{\rm el}} +  \stP{C_{\rm el}}\left(\tfrac{1}{1- \stP{C_{\rm el}}\stP{\nabla U}^T \stP{\mat{H}}^{-1}\stP{\nabla U}} -1\right) \label{eq:c_geom_in_terms_of_c_from_gc_alternate_form}\\
        &=    \stP{C_{\rm el}}\left(\tfrac{1}{1- \stP{C_{\rm el}}\stP{\nabla U}^T \stP{\mat{H}}^{-1}\stP{\nabla U}} \right) = \left(\tfrac{1}{\stP{C_{\rm el}}} - \stP{C_{\rm el}}\stP{\nabla U}^T \stP{\mat{H}}^{-1}\stP{\nabla U} \right)^{-1} \label{eq:c_geom_in_terms_of_c_from_gc}\quad ,
    \end{align}
    we end up with the canonical expression (Eq.~\ref{eq:discussion_gc_geom_cap_first_line}) making use of the triple product (Eq.~\ref{eq:triple_product}) in the first line and using that $\stP{\nabla U}^T \stP{\mat{H}}^{-1}\stP{\nabla U}$ is a scalar and hence:
    \begin{align}
        \stP{\nabla U}^T \stP{\mat{H}}^{-1}\stP{\nabla U}\stP{\nabla U}^T \stP{\mat{H}}^{-1}\stP{\nabla U} = \left(\stP{\nabla U}^T \stP{\mat{H}}^{-1}\stP{\nabla U}\right)^2\quad.\nonumber    
    \end{align}
    While this proofs that both expressions are identical, it still does not provide an intuitive explanation of the apparent contradiction above, that in principle the total capacitance in the canonical case can become negative for a local minimum, while it can not in the grand canonical case. In the following we resolve this apparent contradiction.
    
    \paragraph{Discussion}
    From the derivations above we can make the following statements (again considering a 'normal' electronic response, i.e. $\stP{C_{\rm el}} > 0$):
    \begin{itemize}
        \item[0.] A stationary point in one ensemble is always a stationary point in the other ensemble (but it might change its character).
        The path $\stP{\vec{r}}$ along which the stationary point moves when varying the electrode potential $U$ or the excess charge $q$, respectively, is well defined, it is identical, and it can be computed in in both ensembles (cf. Eq.~\ref{eq:stat_point_in_both_ensembles}).
        \item[1.] The total capacitance (including geometric effects) is identical in the grand canonical and the canonical ensemble (cf. Eq.~\ref{eq:c_geom_in_terms_of_c_from_gc}).
        \item[2.] A local minimum in the grand canonical ensemble:
        \begin{itemize}
            \item [2.1] is always a local minimum in the canonical ensemble (the reverse statement is not necessarily true),
            \item [2.2] it exhibits a positive geometric response (increased total capacitance), i.e. the geometric effect is smaller than the electronic effect in the canonical ensemble ($\stP{\gamma_{\rm geom}}<\stP{\gamma_{\rm el}}$), 
            \item [2.3] and statement 2.2 is true even along every single canonical normal mode of the Hessian.
        \end{itemize}
        \item[3.] A local minimum in the canonical ensemble that fulfills any of the following criteria:
        \begin{itemize}
            \item [3.1] the entire geometric response in the canonical ensemble is larger than the electronic one ($\stP{\gamma_{\rm geom}} > \stP{\gamma_{\rm el}}$),
            \item [3.2] the geometric response along at least one normal mode is larger than the electronic one,
            \item [3.3] the total capacitance is negative,
        \end{itemize}
        is not a local minimum in the grand canonical ensemble and it exhibits a negative total capacitance.
    \end{itemize}
    Note that Statement 3.3 resolves the apparent contradiction, that a local minimum in the canonical ensemble can exhibit a negative total capacitance, while a local minimum in the grand canonical-ensemble always exhibits a positive total capacitance.
    Only local minima with a positive total capacitance are as well local minima in the grand canonical case.
    
    We will justify theses statements in the following.
    We already discussed statements 0. and 1. (see Eqs.~\ref{eq:stat_point_in_both_ensembles} and~\ref{eq:c_geom_in_terms_of_c_from_gc}).
    For showing the other statements, recall the relation between canonical and grand canonical Hessian (Eq.~\ref{eq:gc-Hessian-from-c}):
    \begin{align}
        \mat{\mathcal{H}} = \mat{H} - C_{\rm el} \nabla U \nabla U^T \label{eq:discussion_gcH_from_cH}\\ 
        \Leftrightarrow \mat{H} = \mat{\mathcal{H}} + C_{\rm el} \nabla U \nabla U^T \label{eq:discussion_cH_from_gcH}\quad,
    \end{align} 
    where we used matrix notation.
    
    Keeping in mind that the Hessian defines the character of a stationary point -- for a local minimum the Hessian is positive definite and for a saddle point it is indefinite -- consider now a local minimum $(U, \stP{\vec{r}}(U))$ in the gcPES.
    By definition, the Hessian $\stP{\mat{\mathcal{H}}}$ at this point is positive definite.
    The additional term $\stP{C_{\rm el}} \stP{\nabla U} \stP{\nabla U}^T$ is a positive semi-definite rank one matrix since it is the outer product of two identical vectors, i.e. it has eigenvalues ($\stP{C_{\rm el}} |\stP{\nabla U}|^2, 0,0,...$).
    The canonical Hessian $\stP{\mat{H}}$ hence being the sum of a positive definite and a positive semi-definite matrix therefore is also a positive definite matrix (see Eq.~\ref{eq:discussion_cH_from_gcH}), i.e. the stationary point is a local minimum of the cPES (Statement~2.1).
    
    As discussed already above, the geometric capacitance of a local minimum in the grand canonical ensemble is always positive, due to the fact that $\stP{C_{\rm geom}} = \stP{\nabla q}^T \stP{\mat{\mathcal{H}}}^{-1}\stP{\nabla q}$ is always positive since $\stP{\mat{\mathcal{H}}}$ is positive definite and hence its inverse $\stP{\mat{\mathcal{H}}}^{-1}$ is positive definite as well.
    Similarly, we showed that in the canonical case, the total capacitance of a local minimum can only be positive if the geometric curvature is smaller than the electronic curvature ($\stP{\gamma_{\rm geom}}<\stP{\gamma_{\rm el}}$; cf. Eq.~\ref{eq:geom_curv_c} and statement 4.2 in the canonical \nameref{sec:c_discussion}).
    We can also directly show this mathematically by comparing the geometric capacitance expressed in both ensembles (Eqs.~\ref{eq:discussion_gc_geom_cap} and~\ref{eq:c_geom_in_terms_of_c_from_gc_alternate_form}):
    \begin{align}
        \stP{C_{\rm geom}} &= \stP{C_{\rm total}} -  \stP{C_{\rm el}} \nonumber\\
        = \stP{\nabla q} ^T \stP{\mat{\mathcal{H}}}^{-1} \stP{\nabla q}
        &= \stP{C_{\rm el}} \left(\tfrac{1}{1-\stP{C_{\rm el}} \stP{\nabla U}^T \stP{\mat{H}}^{-1}  \stP{\nabla U}} -1
        \right)\quad \text{i.e.:} \nonumber \\
        \left( \tfrac{\stP{C_{\rm geom}}}{\stP{C_{\rm el}}} = \right)\,\,   \stP{C_{\rm el}}\stP{\nabla U}^T \stP{\mat{\mathcal{H}}}^{-1}\stP{\nabla U}
        &= \left(\tfrac{1}{1-\stP{C_{\rm el}}\stP{\nabla U}^T \stP{\mat{H}}^{-1}\stP{\nabla U}}\right)\quad \stP{C_{\rm el}}\stP{\nabla U}^T \stP{\mat{H}}^{-1}\stP{\nabla U} \quad, \label{eq:geom_cap_over_el_cap_discussion}
    \end{align}
    where we used the triple product to express $\stP{\nabla q}$ in terms of $\stP{\nabla U}$ (Eq.~\ref{eq:triple_product}).
    Note that the term in brackets is in essence the total capacitance expressed in canonical properties (Eq.~\ref{eq:c_geom_in_terms_of_c_from_gc}).
    For a local minimum in the gcPES, the left side of this equation is positive ($\stP{\mat{\mathcal{H}}}$ and $\stP{\mat{\mathcal{H}}}^{-1}$ are positive definite) as well as the second term on the right ($\stP{\mat{H}}$ and $\stP{\mat{H}}^{-1}$ are as well positive definite since a local minimum in the gcPES is a local minimum in the cPES).
    This implies, that the term in brackets, i.e. the total capacitance, has to be positive.
    This can only be the case if $\stP{C_{\rm el}}\stP{\nabla U}^T \stP{\mat{H}}^{-1}\stP{\nabla U} = \tfrac{\stP{\gamma_{\rm geom}}}{\stP{\gamma_{\rm el}}} < 1$.
    This is statement 2.2.
    
    We can go even further by acknowledging again that in the normal mode basis $\{n\}$ of $\stP{\mat{H}}$ we can state that for any normal mode $m$:
    \begin{align}
        1 \stackrel{!}{>} \stP{C_{\rm el}}\stP{\nabla U}^T \stP{\mat{H}}^{-1}\stP{\nabla U}
        &= \stP{C_{\rm el}} \sum_n \tfrac{1}{\stP{H_{n,n}}} \stP{\left(\tfrac{\partial U}{\partial r_n}\right)}^2
        > \stP{C_{\rm el}} \tfrac{1}{\stP{H_{m,m}}} \stP{\left(\tfrac{\partial U}{\partial r_m}\right)}^2 \label{eq:switching_criterion_full}\\
        \Rightarrow \stP{H_{m,m}} &\stackrel{!}{>}  \stP{C_{\rm el}}\stP{\left(\tfrac{\partial U}{\partial r_m}\right)}^2\quad ,\label{eq:switching_criterion_single}
    \end{align}
    since all contributions of the sum are positive.
    Hence, even along every normal mode, the curvature of the purely 'chemical' energy changes caused by an displacement of the system along a normal mode (the eigenvalue of the hessian $\stP{H_{m,m}}$) has to be larger than the curvature of the 'electrostatic' energy changes resulting from the change in potential caused by the displacement ($\stP{C_{\rm el}}\stP{\left(\tfrac{\partial U}{\partial r_m}\right)}^2$) (Statement 2.3).
    
    In reverse, assume we have a local minimum in the canonical ensemble.
    Following the same argumentation, if the geometric response along at least one normal mode is larger than the electronic one ($\stP{H_{m,m}} <  \stP{C_{\rm el}}\stP{\left(\tfrac{\partial U}{\partial r_m}\right)}^2$, case 3.1), then the entire geometric response is larger than the electronic one (case 3.2), and the total capacitance is negative (case 3.3). Hence, the right side of Equation~\ref{eq:geom_cap_over_el_cap_discussion} is negative and the left side has to be as well.
    The grand canonical Hessian therefore can not be positive definite, but must be indefinite.
    The corresponding stationary point in the gcPES is not a local minimum but a saddle point (Statement 3 and comments on Statement 0. and 2.1).
    
    The derived statements about stationary points of the cPES and the gcPES and their relation to each other provide essential information about the influence of geometric effects -- especially when  deriving grand canonical, constant potential results from canonical constant charge simulations.
    While our analysis certainly justifies to follow such a route, i.e. that essentially any quantity discussed can be derived in either ensemble, it also emphasizes that certain differences exist and have to be considered, e.g. the character switching of stationary points.
    In this regard, before we conclude, we want to add a closing note about how the differences behave in the infinite cell size limit.
    
    \paragraph{The infinite cell size limit}
    As a final remark, the criterion $\stP{H_{n,n}} >  \stP{C_{\rm el}}\stP{\left(\tfrac{\partial U}{\partial r_n}\right)}^2$ of a local minimum of the cPES staying a local minimum in the gcPES (Eqs.~\ref{eq:switching_criterion_full} and~\ref{eq:switching_criterion_single}) is essentially given by the ratio between 'chemical' curvature $H_{n,n}$ and 'electrostatic curvature' $\stP{C_{\rm el}}\stP{\left(\tfrac{\partial U}{\partial r_n}\right)}^2$ along a normal mode $n$.
    While the canonical Hessian and its eigenvalues are intensive properties, i.e. system size independent, the electronic capacitance $\stP{C_{\rm el}}$ as is extensive and the gradient of the electrode potential $\stP{\nabla U}$ is inversely extensive.
    This implies that the right side of the criterion in total is inversely extensive, i.e. it becomes smaller with system size.
    For very large systems the criterion is hence always fulfilled and character switching does not occur.
    Similarly, the difference between the canonical and the grand canonical Hessians, which define the character of a stationary point, vanishes in the infinite cell size limit (see Eq.~\ref{eq:gc-Hessian-from-c}).
    From a grand canonical perspective, character switching of stationary points hence can be interpreted as an artifact of canonical, constant charge calculations in small systems: in the infinite cell size limit, where the difference between the canonical and grand canonical ensemble vanishes, corresponding points in both ensembles have identical character.

    \section{Conclusion}
    Based on a detailed mathematical analysis of stationary points in the grand canonical and the canonical ensemble  we derive the geometric response of an electrochemical interface to a change in the applied electrode potential or excess charge, respectively.
    Besides deriving important relationships that allow to compute key properties of the system in either ensemble, our analysis provides a deeper understanding of the geometric response and its contribution to the systems total capacitance.
    As a major result we show, that the character of stationary states in the canonical ensemble might switch when considered in a grand canonical manner -- especially canonical local minima are not necessarily local minima in the grand canonical ensemble.
    We show that such a situation is accompanied by a negative total capacitance of the system and, similarly important, that it is an artifact of small system sizes in canonical 'constant charge' calculations.
    
    This work sets the stage for conducting detailed studies that accurately consider the impact of geometric effects on local minima as well as transition state structures which are crucial for an accurate thermodynamic as well as kinetic description of electrochemical systems.
    
    \begin{acknowledgement}
        The authors gratefully acknowledge funding within the German Research Foundation (DFG) project \mbox{RE1509/33-1} and the DFG CoE e-conversion EXC~2089/1 and support by DFG through the TUM International Graduate School of Science and Engineering (IGSSE).
        Georg Kastlunger acknowledges funding from V-Sustain: The VILLUM Centre for the Science of Sustainable Fuels and Chemicals of VILLUM FONDEN (grant no.~9455).
    \end{acknowledgement}
    
    \clearpage
    
    \bibliography{bibliography}
\end{document}